\documentstyle[12pt,epsf]{article}

\newlength{\pubnumber} 

\catcode`\@=11
\@addtoreset{equation}{section}

\def\section{\@startsection{section}{1}{\z@}{3.5ex plus 1ex minus .2ex}
 {2.3ex plus .2ex}{\large\bf}}
\def\subsection{\@startsection{subsection}{2}{\z@}{2.3ex plus .2ex}
 {2.3ex plus .2ex}{\bf}}

\def\lsim{{\buildrel <\over \sim}}

\def\half{{\textstyle{1\over 2}}}

\def\sixth{{\textstyle {1\over6}}}

\begin{document}

\begin{titlepage}
\samepage{
\setcounter{page}{1}
\rightline{BU-HEPP-02/11}
\rightline{CTP-TAMU-02/04}
\rightline{\tt hep-ph/0303499n}
\rightline{October 2002}
\vfill
\begin{center}
 {\Large \bf String Cosmology: A Review}
\vfill
\vskip .3truecm
\vfill {\large
        Gerald B. Cleaver,\footnote{Gerald{\underline{\phantom{a}}}Cleaver@baylor.edu}}
\vspace{.12in}

{\it        Center for Astrophysics, Space Physics \& Engineering Research\\
            Department of Physics, PO Box 97316, Baylor University\\
            Waco, Texas 97316, USA\\
            and\\
            Astro Particle Physics Group,
            Houston Advanced Research Center (HARC),\\
            The Mitchell Campus,
            Woodlands, TX 77381, USA\\}
\vspace{.025in}
\end{center}
\vfill
\begin{abstract}
The second string revolution, which begin around 1995, has led to a drastic
alteration in our perception of the universe, perhaps even more so then did 
the first string revolution of 1984. That is, extending 10-dimensional 
string theory to 11-dimensional $M$-theory has had more profound implications 
than did the original extension of 4-dimensional quantum mechanics and 
relativity to 10-dimensional string theory. After a brief review of $M$-theory,
I discuss some implications of large extra dimensions. I then consider 
astronomical evidence for, and constraints on, large compactified dimensions. 
I conclude with a possible resolution to the apparent inconsistency between 
the MSSM scale and string scale in the weak coupling limit.
\end{abstract}
\begin{center}
{\it Talk presented at COSPAR '02, Houston, Texas, October 2002.}
\end{center}
\smallskip}
\end{titlepage}

\setcounter{footnote}{0}

\section{Cosmology and the Second String Revolution}

The first string revolution began place in 1984 after John H.\ Schwarz and
Michael Green demonstrated anomaly cancellation in the Type I superstring
(Green 1984). Similar anomaly cancellation was soon shown, thereafter, for 
the four other superstring theories, Type IIA, Type IIB, Heterotic $SO(32)$ and 
Heterotic $E_8\times E_8$. Superstring theory was quickly recognized as the
leading candidate for a ``Theory of Everything.'' String theory
unified the four known forces and explained all particles as different modes
of a fundamental ``string''. However, it also drastically transformed our 
understanding of the universe, revealing that it
more complex than had ever been imagined prior. 
For no longer was our universe $(3+1)$-dimensional,
instead it became (9+1)-dimensional. The six new dimensions were claimed to 
necessarily be compactified with Planck-scale length to explain the matter 
and non-gravitational force content of our universe.

The problem with string theory was that it wasn't a single theory, but actually
five. For the next decade one of the primary
directions in string research was investigation of the parameter space of 
each of these five theories and of connections between the theories. 
While each theory contained only one
solution with ten large non-compact dimensions,
each time another dimension was
compactified the number of solutions grew ever more enormous. 

Soon after the first revolution, equivalences were being recognized between 
different models. The simplest involved models 
with only one compactified dimension: 
replacing a compactification radius $R_1$ in one model with
with radius $R_2 = \alpha^{'}/R_1$, 
where $\alpha^{'}\equiv {1\over 2\pi T}$ is the 
string slope (with T the string tension)
produces another physically equivalent model. 
In the watershed year of 1995,
Ed Witten, Joe Polchinski, Petr Horova and colleagues demonstrated 
equivalences between models of different 
string theories (Horava 1996, Polchinski 1996) 
and even related 10-dimensional string theory to an eleven dimensional
supergravity theory. Thus was born the second string revolution.
It was soon understood that all five 10-dimensional string theories and
11 dimension supergravity where all part of one more encompassing theory.
Each of the old theories were different regions 
in the parameter space of a single 11-dimensional theory, now known as $M$-theory. 
In string language,
the size of the eleventh dimension was found to corresponded to the 
string coupling strength. Relatedly, the eleventh dimension also gave strings
thickness, thus transforming them into membranes.
In models with weak coupling the 11$^{\rm th}$-dimension is of Planck size 
or smaller, while in strong coupled string models it 
is larger than the Planck scale. Thus, the eleventh 
dimension of spacetime becomes apparent in a ten-dimensional string theory
as the string coupling strength becomes strong. 

The idea of the $11^{\rm th}$ dimension growing large led several groups
to consider allowing varying numbers of the other six compactified dimensions to also grow large.
Two of the first groups were Arkani-Dimopoulos-Dvali (ADD) and Randall-Sundrum (RS)
The former group developed models where only gravity propagates in the large extra dimensions
(Arkani-Hamed 1998, Antoniadis 1998).
In contrast, the latter group allowed varying combinations of matter and vector boson fields
to propagate in all dimensions (Randall 1999). 
The ADD model allows many dimensions to be much larger than the 
Planck scale, while the RS model allows only the 11$^{\rm th}$ dimension to grow large. 

When gravity propagates in the large extra dimensions  
the classical gravitational force law goes as 
$\textstyle 1\over r^{2+n}$ (derived from Gauss' Law in $3+n$ dimensions), 
where $3+n$ is the number of non-compact and compact dimensions 
 of size significantly greater than $r$. If these $n$ extra dimensions are of the ADD class,
then the standard model forces and gravity unify at an energy scale $M^{\ast}$ if the size 
of these dimensions are all of the scale,  
$R^{\ast}\equiv {\hbar c\over M^{\ast} c^2}\left( {M_P\over M^{\ast} } \right)^{(2/n)}$.
In this manner, unification of all forces 
is possible at a scale as low as a few TeV. This would imply, though, that at least
one dimension is of sub-millimeter length. 
Unification occurs because the gravitational coupling strength grows larger at higher energy scales,
increasingly faster as $n$ increases. Although $n=1$ has been ruled out by 
astrophysical constraints, $n=2$ yields $R^{\ast}\sim .1$ mm.

String/M-theory considerations can modify the inverse $r^{2+n}$ form of the force law 
to a more general expression with potential    
$V(r) = -G {m_1 m_2\over r} ( 1 + \alpha {\rm e}^{-r/\lambda} )$. Two large extra dimensions
yields $\lambda = R^{\ast}$ and $\alpha = 3$ $(4)$ for compactification on a 2-sphere
(2-torus). Dilaton and moduli exchange correspond to $\alpha$ as large as $10^5$ and
$\lambda\sim 0.1$ mm.

While the unification of forces via dimensions of sub-millimeter scale  
eliminates the hierarchy between the weak and the Planck distance scales,
it creates another: that between the weak scale and the much larger compactification scale $R^{\ast}$.     
In their model Randall-Sundrum sought to avoid introducing this new hierarchy, so they allowed 
only the eleventh dimension to grow beyond Planck scale and only by a factor of 
$M_{\rm Plank}/M_{\rm GUT}$. Near the Planck scale, spacetime appears 5-dimensional, but it's metric
is ``warped'': $ds^2 = {\rm e}^{-2k r_c \phi}\eta_{\mu\nu} dx^{\mu} dx^{\nu} + r^2_c d\phi^2$,
where $k$ is of Planck scale order, $x^{\mu}$ are coordinate for our four dimensions, and
$0 \leq \phi \leq \pi$ is the coordinate of the $11^{\rm th}$ dimension, for which 
$r_c$ is the size scale. This metric was shown to be a solution to Einstein's equations with two
3-branes (tensored with six compactified dimensions) separated along the direction of $\phi$.     
In this model the large hierarchy between the Planck and weak scales does not require $r_c$ to be
extremely large compared to the Planck scale, but only larger by a factor of around 50. 
The hierarchy is generated by the exponential term    
that decreases in values as $\phi$ increases. This corresponds to movement away from one 3-brane 
(a hidden universe) towards the other 3-brane (our universe).    

\subsection{Varying Physical Constants}

In string/M-theory the traditional physical constants in nature, 
such as the speed of light,
are not fundamental constants. Their values depend upon the geometry of
the extra spatial dimensions, specifically their size and arrangement. Thus, 
these physical ``constants'' may be dynamical during different phases of an 
evolving string universe. 
The sizes of the compactified dimensions correspond to scalar (moduli) fields of the theory. 

Recent astronomical evidence suggests that at least one of the extra 
spatial dimensions, in particular the $11^{\rm th}$, in string theory might indeed be very large.
The first piece of evidence originates in  magnesium and iron atom absorption lines in
the light emitted by quasars billions of light years away. The absorption lines are created
as the emitted light passes through interstellar gas. The positions of the lines
are functions of the fine-structure constant.  Based on their analysis of 
light from over 30 distant quasars that contain these absorption lines, Jack Webb 
et al. have concluded that the fine-structure constant, $\alpha_{em}$, has been varying 
significantly only a few billion years ago (Webb 1999, 2000). 
When the wavelengths for
specific quasar absorption lines was compared to those in absorption lines produced in a
laboratory, a consistent shift in the quasar lines from four independent sets of data 
was found suggesting that a few billion years ago 
$\alpha_{em}\equiv \textstyle{e^2\over \hbar c}$ was about 
smaller than it is now. Comparison of the lab value of $\alpha$ to the value of $\alpha$ 
from quasars within the red shift domain  $z\sim 0.6$ to 2 indicated a $4 \sigma$ deviation, 
$\delta \alpha/ \alpha = (0.72\pm 0.18)\times 10^{-5}$.
This implies $e$ has increased and/or $c$ has decreased over the 
last few billion years.  Changes in $c$ in particular suggest a low energy M-theory 
scale and a related long time before some compactified dimensions stabilized. 

P.C.W.\ Davies et al.\ have recently argued that black hole thermodynamics suggests that 
only $c$ changed (Davies 2002). Otherwise, they claim that an increase in $e$ would have decreased the
entropy of black holes, violating the generalized second law of thermodynamics. However,
S.\ Carlip and S.\ Vaidya have contested this claim, arguing that when the full thermal
environment of a black hole is considered, thermodynamics is consistent with a decrease in $c$,
an increase in $e$, or both (Carlip 2002). Further, for a certain class of charged dilaton black holes 
related to string theory, M.\ Fairbairn and M.\ Tytgat have shown that the entropy does not change under 
adiabatic variations of $\alpha$ and one might expect it to increase for non-adiabatic changes 
(Fairbairn 2002).

M. Duff has also argued that, while the possible time variation 
of dimensionless fundamental constants of nature, such as the fine structure constant $\alpha$, is a 
legitimate subject of physical enquiry, the time variation of dimensional constants, such as 
$\hbar$, $c$, $G$, $e$, $k$..., has no operational meaning, since the latter are merely human constructs 
whose number and values differ from one choice of units to the next (Duff 2002).

P.\ Brax et al.\ have investigated under 
what conditions masses or gravitational constant or any of the three 
Standard Model coupling constants, including $\alpha$, might vary (Brax 2002). 
They determined that
in the brane world scenario, an evolution of masses or of the gravitational constant (which one 
depends on the frame) is generally predicted when the moduli fields are not stabilized, whereas
an evolution of a coupling constant is predicted only under more special circumstances 
(Brax 2002).
Only if vector bosons are directly coupled to the bulk scalar field in a 
string/M-theory model, which is often the case,
would the corresponding coupling constants vary. 
Thus, a time-variation of coupling constants is a prediction of 
string/M-theory (Brax 2002).

\subsection{The GZK Paradox}

A current astrophysical paradox is the detection on earth of ultra high energy cosmic rays
with energies beyond the Greisen-Zatsepin-Kuzmin (GZK) cutoff 
(Greisen 1966, Zatsepin 1966). There are
no known astrophysical sources in the local region of our galaxy that can account for these
streams of particles, which appear to be hadrons, rather than photons or neutrinos, based on
the signatures of shower events as these high energy particles interact with our atmosphere.
However, if the particles are truly hadronic in nature, then their propagation over astrophysical
distances is strongly influenced by the cosmic background radiation. These interactions result 
in the GZK cutoff on the maximum energy of cosmic ray hadrons, with 
$E_{\rm MAX}\le 10^{20}$ eV. Further, if the particles were photonic, rather than hadronic,
then they would be expected to have a mean free path of less than 10 Mpc due to scatter with the 
cosmic background radiation and radio photons (Coriano 2002). Therefore, unless the cosmic ray
particles are neutrinos, their origin must be nearby. However, neutrinos would interact very weakly
in our atmosphere, not producing the signatures seen. 

String/M-theory offers at least two distinct solutions to the GZK paradox. 
The first proposal is that the ultra high energy cosmic rays originate
from the decay of long-lived super-heavy matter states with mass of order $10^{12-15}$ GeV, which
simultaneously are good candidates for cold dark matter (Coriano 2001, 2002). 
Then these cosmic rays
could have their origin in decays within our galactic halo and escape GZK bounds (Coriano 2002).
Matter with the basic properties required to produce such cosmic rays are inherent
to a class of (semi)-realistic models derived from the heterotic string and
originate from Wilson line breaking of GUT symmetries.   
More specifically, though, the super-heavy states must have a lifetime of order $10^{17}$ seconds to
$10^{28}$ seconds and their abundance should satisfy the relation, 
$(\Omega_X / \Omega_O) (t_0/\tau_X)\sim 5\times 10^{-11}$ to account for the observed flux of cosmic
rays through the earth's atmosphere. $t_0$ is the age of the universe,  $\tau_X$ is the lifetime 
of the meta-stable state, $\Omega_0$ is the critical mass density, and $\Omega_X$ is the relic 
mass density of the meta-stable state (Coriano 2002). Whether or not these latter properties are 
specifically satisfied is very string model dependent. Potential ultra high energy
ray producing candidates include ``cryptons'' in the flipped $SU(5)$ model (Antoniadis, Lopez), 
which are condensates of $\bf 4$ and $\bar{\bf 4}$ reps of the hidden sector $SU(4)_H$, 
``unitons'', which are exotic standard model quarks and carry a fractional $U(1)_{Z'}$ charge,
and ``singletons'', which is a standard model singlets but carry a fractional $U(1)_{Z'}$ charge   
(Coriano 2001).

An alternate or complimentary explanation for the GZK paradox is based upon M-theory
space-time foam effects studied by Ellis et al. They model spacetime foam using 
a non-critical Liouville-string model for the quantum fluctuations of $D$-branes with recoil.
Ellis et al.\  argue that particle momentum could be conserved exactly during propagation
but only on the average during interactions with a $D$-brane, while energy is conserved only
on the average during interactions and, in general, is not conserved during interactions with
a brane because of changes in the background metric. Ellis et al.\ conclude that
$D$-brane recoil effects provide another means by which the GZK cut-off can be avoided (Ellis 2001).

\section{Constraints on Sizes of Extra Dimensions}
 
The scale below which 
deviation from $1/r^2$ begins corresponds to the size of the largest compactified string 
or M-theory dimension. 
Hoyle et al.\ at the University of Washington have 
verified from a Cavendish torsion-type experiment
that gravity keeps its $1/r^2$ form down to 0.218 mm. 
This translates into a 95$\%$ confidence upper limit of .150 mm.\ on the size of two compact
dimensions and a 95$\%$ confidence that the largest compact dimension of any number is less 
than .2 mm.\ (Hoyle 2000, Adelberger 2002).
 
Even stronger constraints have been imposed from neutrino flux measurements of the
SN 1987a supernova. Hanhart et al.\  have developed self-consistent simulations of the early,
neutrino-emitting phase of a proto-neutron star which include energy losses due to the coupling
of Kaluza-Klein modes of a graviton which arise with ADD compactified dimensions. 
They compared the neutrino signals from their simulations to that from SN 1987a and from a
probabilistic analysis determined the upper bound for two compact extra dimensions to be 
0.66 $\mu$m at the $95\%$ confidence level and, similarly, 
an upper bound of 0.8 nm.\ for three extra dimensions (Hanhart 2001).

Milton has placed not upper, but lower limits on large compactified dimensions.
Since quantum fields in extra compact dimensions should give rise to a quantum
vacuum or Casimir energy and supernova and cosmic microwave background data indicate 
that the cosmological constant is of the same order as the critical mass density of the universe, 
Milton argues a lower bound of around 10 $\mu$m.\ to the size of large compact dimensions (Milton 2000). 
Otherwise he claims the Casimir energy would produce too large of a cosmological constant. 

\section{MSSM and String Scales Consistency}

Strong coupling effects of $M$-theory can lower (Witten)
heterotic string scale $\Lambda_H$ down to $\Lambda_U$.
However, when a weak string coupling is assumed,
that is, $R_{11}\lsim 10^{-33}$ cm.,
an enduring issue has been the discrepancy between 
the $SU(3)_C\times SU(2)_L\times U(1)_Y$ ([321])
gauge coupling unification scale,
$\Lambda_U \approx 2.5 \times 10^{16}$ GeV (Ellis 1990, Amaldi 1991, Langacker 1991),
for the the MSSM with intermediate scale desert
and the string scale, $\Lambda_H \approx 5\times 10^{17}$ GeV (Kaplunovsky 1988),
for the weakly coupled heterotic string.

Two weak coupling solutions have been proposed to
resolve this factor of 20 disagreement (Dienes 1995a, 1995b, 1997).
One proposal is a grand unified 
theory between $\Lambda_U$ and $\Lambda_H$. Here the MSSM couplings
merge at $\Lambda_U$ and then run together within a GUT to
$\Lambda_H$.
However, with the exception of flipped $SU(5)$ (Antoniadis 1989, Lopez 1993)
(or partial GUTs such as the Pati-Salam $SU(4)_C\times SU(2)_L\times SU(2)_R$
(Pati 1975, Chang 1984)
string GUT models based on level-one Ka\v c-Moody algebras encounter a
difficulty: they lack the required adjoint higgs (and
higher representations).
Alternately, intermediate scale exotics could shift
the MSSM unification scale upward to the string scale (Chang 1997).

The near ubiquitous appearance of MSSM-charged exotics in
heterotic string models adds weight to this third proposal.
If MSSM exotics exist with
intermediate scale masses of order $\Lambda_I$, then the actual
[321] running couplings are altered above $\Lambda_I$. It is then,
perhaps, puzzling that the illusion of MSSM unification
should still be maintained when the
intermediate scale MSSM exotics are ignored (Giedt 2002).
Maintaining this illusion
likely requires very fine tuning of $\Lambda_I$ for a generic
exotic particle set and $\Lambda_H$.
Slight shifting of $\Lambda_I$ would, with high probability,
destroy appearances.
Thus, in some sense, the apparent MSSM unification below the
string scale might be viewed as accidental (Ghilencea 1999, Giedt 2002).

A mechanism whereby the appearance of
a $\Lambda_U$ is not accidental would be very appealing.
Just such a mechanism, entitled ``optical unification,''
has recently been discussed by J.\ Giedt (Giedt 2002).
Optical unification results in $\Lambda_U$ not disappearing
under shifts of $\Lambda_I$.
Instead, $\Lambda_U$ likewise shifts in value.
This effect is parallel to a virtual image always appearing
between a diverging lens and a real object,
independent of the position of the lens or real object.
Hence, Giedt's choice of appellation for this mechanism.

Successful optical unification requires three things (Giedt 2002).
First, the effective level of the
hypercharge generator must be the standard
$k_Y = \textstyle{5\over3}$.
This is a
strong constraint on string-derived $[321]$ models,
for the vast majority have non-standard
hypercharge levels.
Only select classes of models, such as the NAHE-based
free fermionic class,
can yield $k_{Y} = \textstyle{5\over3}$.
Second, optical unification imposes the relationship
$\delta b_2 = \textstyle{7\over12} \delta b_3 + \textstyle{1\over4}
\delta b_Y$,
between the exotic particle contributions $\delta b_3$, $\delta b_2$,
and $\delta b_1$
to the [321] beta function coefficients.
Each $SU(3)_C$ exotic triplet or anti-triplet contributes
$\half$ to $\delta b_3$;
each $SU(2)_C$ doublet contributes
$\half$ to $\delta b_2$.
With the hypercharge of a MSSM quark doublet normalized to $\sixth$,
the contribution to $\delta b_Y$ from an individual particle with
hypercharge $Q_Y$ is $Q_Y^2$.
$\delta b_3 > \delta b_2$ is required
to keep the virtual unification scale below the string scale.
Combining this with the second constraint imposes
$\delta b_3 >  \delta b_2 \ge \textstyle{7\over12} \delta b_3$,
since $\delta b_Y \geq 0$.

To acquire intermediate scale mass,
the exotic triplets and anti-triplets must be equal in number.
Similarly, the exotic doublets must be even in number.
Hence, $\delta b_3$ and $\delta b_2$ must be integer (Giedt 2002).
As Giedt pointed out, the simplest solution to optical unification
three exotic triplet/anti-triplet pairs and two pairs of doublets.
One pair of doublets can carry $Q_Y=\pm \half$, while the remaining
exotics carry no hypercharge.
Alternately, if the doublets carry too little hypercharge,
some exotic $SU(3)_C \times SU(2)_L$ singlets could make up the
hypercharge deficit.
The next simplest
solution requires four triplet/anti-triplet pairs and three pairs of
doublets that yield $\delta b_Y = 2 \textstyle{2\over 3}$
either as a set, or with the assistance of additional non-Abelian singlets.
Cleaver et al.\  presents a standard-like model that has the potential
to realize the latter optical unification solution,
with the required hypercharge carried by the (anti)-triplets and
one additional pair of singlets (Cleaver 2002).
Detailed analysis of this model is underway.

\section{Concluding Comments}

The second string revolution answered many of the fundamental questions
of string theory, in particular, the relation between the five former
string theories. But M-theory has also raised more 
questions than it has answered. What M-theory truly is, is not known. 
String theory, even with ten-dimensions, 
presented a much less complicated picture of the universe 
than $M$-theory does. 
The drastically different string cosmologies that the ADD and RS
theories present give a hint to the vast array of possibilities 
allowed perturbatively by $M$-theory. For now we can only wonder 
about the form of the true $M$-theory universe.

\section*{ACKNOWLEDGEMENTS}

In this talk I have only scratched the surface of recent developments
in string/$M$-theory research and related cosmological implications. 
The work of many, many other people should be mentioned 
in a more complete review of recent progress in string/$M$ cosmology.
I thank Michel Tytgat for bringing some manuscripts to my attention. 

\section*{REFERENCES}

{\def\nh{\noindent\hangindent=1pc\hangafter=1  }

\nh Adelberger E.G., et al., 
Sub-millimeter tests of the gravitational inverse square law, 2002, [hep-ex/0202008].

\nh Amaldi, U., W.\  de Boer and F.\  F\"urstenau, 
Comparison of grand unified theories with electroweak and strong coupling constants 
measured at LEP, {\it Phys.\ Lett.}, {\bf B260}, 447, 1991.

\nh Antoniadis, I., J.\ Ellis, J.\ Hagelin, and D.V.\ Nanopoulos,
The flipped $SU(5)$ model revamped, {\it Phys.\ Lett.}, {\bf B231}, 65, 1989.

\nh Antoniadis, I., N.\ Arkani-Hamed, S.\ Dimopoulos, and G.\ Dvali, New dimensions
at a millimeter to a fermi and superstrings at a TeV, {\it Phys.\ Lett.}, {\bf B436},
257, 1998, [hep-ph/9804398]. 

\nh Arkani-Hamed, N., S.\ Dimopoulos, and G.\ Dvali, The hierarchy problem and new dimensions
in millimeters, {\it Phys.\ Lett.}, {\bf B429}, 263, 1998, [hep-ph/9803315]. 

\nh Brax, P., C.\  van de Bruck, A.C. Davis, et al., Varying constants in Brane World
Scenario, {\it Astrophys.\ Space Sci.}, {\bf 283}, 627, 2003, [hep-ph/0210057].

\nh Carlip, S., and S.\  Vaidya, Black holes may not constrain varying constants,
{\it Nature}, {\bf 421}, 498, 2003, [hep-th/0209249]; 
Carlip, S., Varying constants, black holes and quantum gravity, 
{\it Phys.\ Rev.}, {\bf D67}, 023507, 2003, [gr-qc/0209014].

\nh Chang, D., R.N.\  Mohapatra, A new approach to left-right symmetry breaking in 
unified gauge theories, {\it Phys.\ Rev.} {\bf D30}, 1052, 1984.

\nh Chang, S., C. Coriano, and A. Faraggi, New dark matter candidates motivated from 
superstring derived unification, {\it Phys.\ Lett.}, {\bf B397}, 76, 1997,
[hep-ph/9603272].

\nh Cleaver, G., V.\ Desai, H.\ Hanson, et al.,
On the Possibility of Optical Unification in Heterotic Strings,
{\it Phys.\ Rev.}, {\bf D67}, 026009, 2003, [hep-ph/0209050].

\nh Coriano, C., A.E.\ Faraggi, and M.\ Pl\" umacher, Stable Superstring Relics and 
Ultrahigh Energy Cosmic Rays, {\it Nucl.\ Phys.} {\bf B614}, 2001, [hep-ph/0107053].

\nh Coriano, C., and A.E.\ Faraggi, Seeking Experimental Probes of String Unification, 2002,
[hep-ph/0201129].

\nh Davies, P.C.W., T.M.\ Davis, and C.H. Linewater, Black holes constrain varying constants,
{\it Nature}, {\bf 418}, 602, 2002.

\nh Dienes, K.R., and A.E. Faraggi, Making ends meet, {\it Phys.\ Rev.\ Lett.}, {\bf 75}, 2646, 1995a, [hep-th/9505018].

\nh Dienes, K.R., and A.E. Faraggi, Gauge coupling unification in realistic free fermionic string models, 
{\it Nucl.\ Phys.}, {\bf B457}, 409, 1995b, [hep-th/9505046].

\nh Dienes, K.R., {\it Nucl.\ Phys.} {\bf B487}, 447, 1997, [hep-th/9602045],
and references therein.

\nh Duff, M., Comment on time-variation of physical constants, MCTP-02-43, 2002,
[hep-th/0208093].

\nh Ellis, J.,  S.\ Kelley, and D.V. Nanopoulos, Precision LEP data, supersymmetric guts 
and string unification, {\it Phys.\ Lett.}, {\bf B249}, 441, 1990.

\nh Ellis, J., N.E.\ Mavromatos, and D.V. Nanopoulos, How large are dissipative effects in noncritical 
Liouville string theory? {\it Phys.\ Rev.} {\bf D63}, 024024, 2001.

\nh Fairbairn, M., and M.\ Tytgat, Varying alpha and black hole entropy, {\it JHEP}, {\bf 0302}, 005, 2003, 
[hep-th/0212105].

\nh Ghilencea, D., Predictions for alpha$_{3_{M_{Z}}}$ in a string inspired model, 
{\it Phys.\ Lett.}, {\bf B459}, 540, 1999, [hep-ph/9904293]

\nh Giedt, J., Optical Unification, {\it Mod.\ Phys.\ Lett.}, {\bf A18}, 1625, 2003,
[hep-ph/0204315].

\nh Green, M., J.H. Schwarz, Anomaly cancellation in supersymmetric $D=10$ gauge theory and 
superstring theory, {\it Phys.\ Lett.}, {\bf B149}, 1007, 1984.

\nh Greisen, K., End to the cosmic ray spectrum? 
{\it Phys.\ Rev.\ Lett.}, {\bf 16}, 748, 1966.

\nh Hanhart, C., J.\ Pons, D.\ Phillips, S.\ Reddy, The likelihood of gods' existence:
improving the SN 1987a constraint on the size of large compact dimensions, 
{\it Phys.\ Lett.}, {\bf B509}, 1, 2001, [astro-ph/0102063].

\nh Horava, P., and E.\ Witten, Heterotic and type I string dynamics 
from eleven dimensions, {\it Nucl.\ Phys.} {\bf B460}, 506, 1996.

\nh Hoyle, C.D., U.\ Schmidt, B.R.\ Heckel et al. Sub-millimeter test of the gravitational 
inverse-square law: a search for ``large'' extra dimensions, {\it Phys.\ Re.\ Lett.},
86, 1418, 2001,  [hep-ph/0011014].

\nh Kaplunovsky, V.S., One loop threshold effects in string unification,
{\it Nucl.\ Phys.} {\bf B307} 145, 1988; Erratum
{\it Nucl.\ Phys.}, {\bf B382}, 436, 1992.

\nh Langacker, L. and M. Luo, Implications of precision electroweak experiments for $M_T$, 
rho$_o$, sin$^2 \theta_w$ and grand unification, {\it Phys.\ Rev.}, {\bf 44}, 817, 1991.

\nh Lopez, J, D.V.\ Nanopoulos and K.\ Yuan, 
The search for a realistic flipped $SU(5)$ string model, 
{\it Nucl.\ Phys.}, {\bf B399}, 654, 1993.

\nh Milton, K., Constraints on extra dimensions from cosmological and terrestrial measurements,
{\it Gravitation \& Cosmology}, {\bf 6}, 1-10, 2000, [hep-th/0107241].

\nh Pati, J. and R.N. Mohapatra, A natural left-right symmetry, 
{\it Phys.\ Rev.}, {\bf D11}, 2558, 1975. 

\nh Polchinski, J., and E.\ Witten, Evidence for heterotic-type I string 
duality, {\it Nucl.\ Phys.} {\bf B460}, 525, 1996. 

\nh Randall, L.\  and R.\ Sundrum, A large mass hierarchy from a small extra dimension,
{\it Phys.\ Rev.\ Lett.}, {\bf 83}, 3370, 1999. 

\nh Webb, J.K., V.V.\  Flambaum, C.W.\  Churchill, et al. Search for time variation of the fine 
structure constant, {\it Phys.\  Rev.\  Lett.}, {\bf 82}, 884-887, 1999.

\nh Webb, J.K., V.V.\  Flambaum, C.W.\  Churchill, et al., Further evidence for cosmological 
evolution of the fine structure constant. {\it Phys.\ Rev.\ Lett.}, {\bf 87}, 091301, 2001.

\nh Witten, E., Strong coupling expansion of calabi-yau compactification, 
{\it Nucl.\ Phys.}, {\bf B471}, 135, 1996, [hep-th/9602070].

\nh Zatsepin, G.T., and V.A.\ Kuzmin, Upper limits of the spectrum of cosmic rays,
{\it Pisma Zh.\ Eksp.\ Theor.\ Fiz.} {\bf 4}, 114, 1966.

}


\end{document}